\documentclass{jfm}

\usepackage{graphicx}
\usepackage{newtxtext}
\usepackage{newtxmath}
\usepackage{natbib}
\usepackage{hyperref}
\hypersetup{
    colorlinks = true,
    urlcolor   = blue,
    citecolor  = black,
}

\newcommand{\RomanNumeralCaps}[1]

\usepackage[usenames,dvipsnames,svgnames,table]{xcolor}
\usepackage[capitalise]{cleveref}
\usepackage{graphicx}
\usepackage{dcolumn}
\usepackage{bm}
\usepackage{xcolor}
\usepackage{verbatim} 
\usepackage{float}

\newcommand{\gv}[1]{\ensuremath{\mbox{\boldmath$ #1 $}}} 
\newcommand{\bv}[1]{\ensuremath{\boldsymbol{#1}}} 

\newcommand{\dev}{'}
\newcommand{\Ca}{\mathrm{Cau}}
\newcommand{\order}[1]{\mathcal{O} \left( #1 \right)}

\renewcommand{\cite}[1]{\citep{#1}}

\title{Three-dimensional soft streaming}

\author{Songyuan Cui\aff{1},
        Yashraj Bhosale\aff{1},
        \and
        Mattia Gazzola\aff{1}\aff{2}\aff{3} \corresp{\email{mgazzola@illinois.edu}}
 }

\affiliation{
\aff{1}Mechanical Sciences and Engineering, University of Illinois at Urbana-Champaign, Urbana, IL 61801, USA
\aff{2}National Center for Supercomputing Applications, University of Illinois at Urbana-Champaign, Urbana, IL 61801, USA
\aff{3}Carl R. Woese Institute for Genomic Biology, University of Illinois at Urbana-Champaign, Urbana, IL 61801, USA
}

\begin{document}
\maketitle

\begin{abstract}
Viscous streaming is an efficient rectification mechanism to exploit flow inertia at small scales for fluid and particle manipulation. It typically entails a fluid vibrating around an immersed solid feature that, by concentrating stresses, modulates the emergence of steady flows of useful topology. Motivated by its relevance in biological and artificial settings characterized by soft materials, recent studies have theoretically elucidated, in two dimensions, the impact of body elasticity on streaming flows. Here, we generalize those findings to three dimensions, via the minimal case of an immersed soft sphere. We first improve existing solutions for the rigid sphere limit, by considering previously unaccounted terms. We then enable body compliance, exposing a three-dimensional, elastic streaming process available even in Stokes flows. Such effect, consistent with two-dimensional analyses but analytically distinct, is validated against direct numerical simulations and shown to translate to bodies of complex geometry and topology, paving the way for advanced forms of flow control.

\end{abstract}


\section{Introduction}\label{sec:intro}
This study investigates the effects of body elasticity on three-dimensional viscous streaming. 
Viscous streaming, an inertial phenomenon, refers to the steady, rectified flows that emerge when a fluid oscillates around a localized microfeature. 
Given its ability to remodel surrounding flows over short time and length scales, streaming has found application in multiple aspects of microfluidics, from particle manipulation \cite{lutz2003microfluidics,lutz2005microscopic,marmottant2004bubble,lutz2006hydrodynamic,wang2011size,chong2013inertial,chen2014manipulation,klotsa2015propulsion,thameem2017fast,pommella2021enhancing} and chemical mixing~\cite{liu2002bubble,lutz2003microfluidics,lutz2005microscopic,ahmed2009fast} to vesicle transport and lysis~\cite{marmottant2003controlled,marmottant2004bubble}. 
Recently, the use of multi-curvature streaming bodies has expanded the ability to manipulate flows, leading to compact, robust, and tunable devices for filtering and separating both synthetic and biological particles \cite{parthasarathy2019streaming,bhosale_parthasarathy_gazzola_2020,chan2021three,bhosale2022multicurvature}.
More recently yet, motivated by medical and biological applications, the effect of body compliance has been considered \cite{pande2023oscillatory,anand2020transient}, with one of the studies yielding a first two-dimensional streaming theory for soft cylinders \cite{bhosale2022streaming}. 
Its major outcome is encapsulated in the relation
\begin{equation}
\label{eqn:final_result_2d}
    \langle \psi_1 \rangle = \sin 2 \theta ~ \left[ \Theta(r) + \Lambda(r) \right]
\end{equation}
where $\langle \psi_1 \rangle$ is the time-averaged Stokes streamfunction, $r$ and $\theta$ are the radial and polar coordinates in the cylindrical system. 
This relation reveals an additional streaming process $\Lambda(r)$, purely induced by body elasticity, that is available even in Stokes flows where rigid-body streaming $\Theta(r)$ cannot exist \cite{holtsmark1954boundary}. 
Elasticity modulation has then been shown to achieve streaming configurations similar to rigid bodies, but at significantly lower frequencies. 
This frequency reduction has relevant implications, as it renders viscous streaming accessible within the limits of biological actuation.

In this work, we extend this understanding to three dimensions by examining the minimal case of an oscillating, soft sphere. 
We first present an improved theoretical solution for the rigid sphere case by augmenting \citet{lane1955acoustical}'s derivation with a previously unaccounted term related to vortex stretching. 
Our formulation is shown to significantly enhance quantitative agreement with direct numerical simulations and experiments.
Next, within the same theoretical framework, we consider body elasticity and seek a modified streaming solution dependent on material compliance. 
We recover an independent elastic modification term, similar in nature to the above two-dimensional result but analytically distinct.
We then demonstrate the accuracy of our theory against direct numerical simulations, and show how observed elasticity effects may translate to bodies of complex geometry and topology, further expanding the potential utility of soft streaming.

\begin{figure}
\centering 
\includegraphics[width=\linewidth]{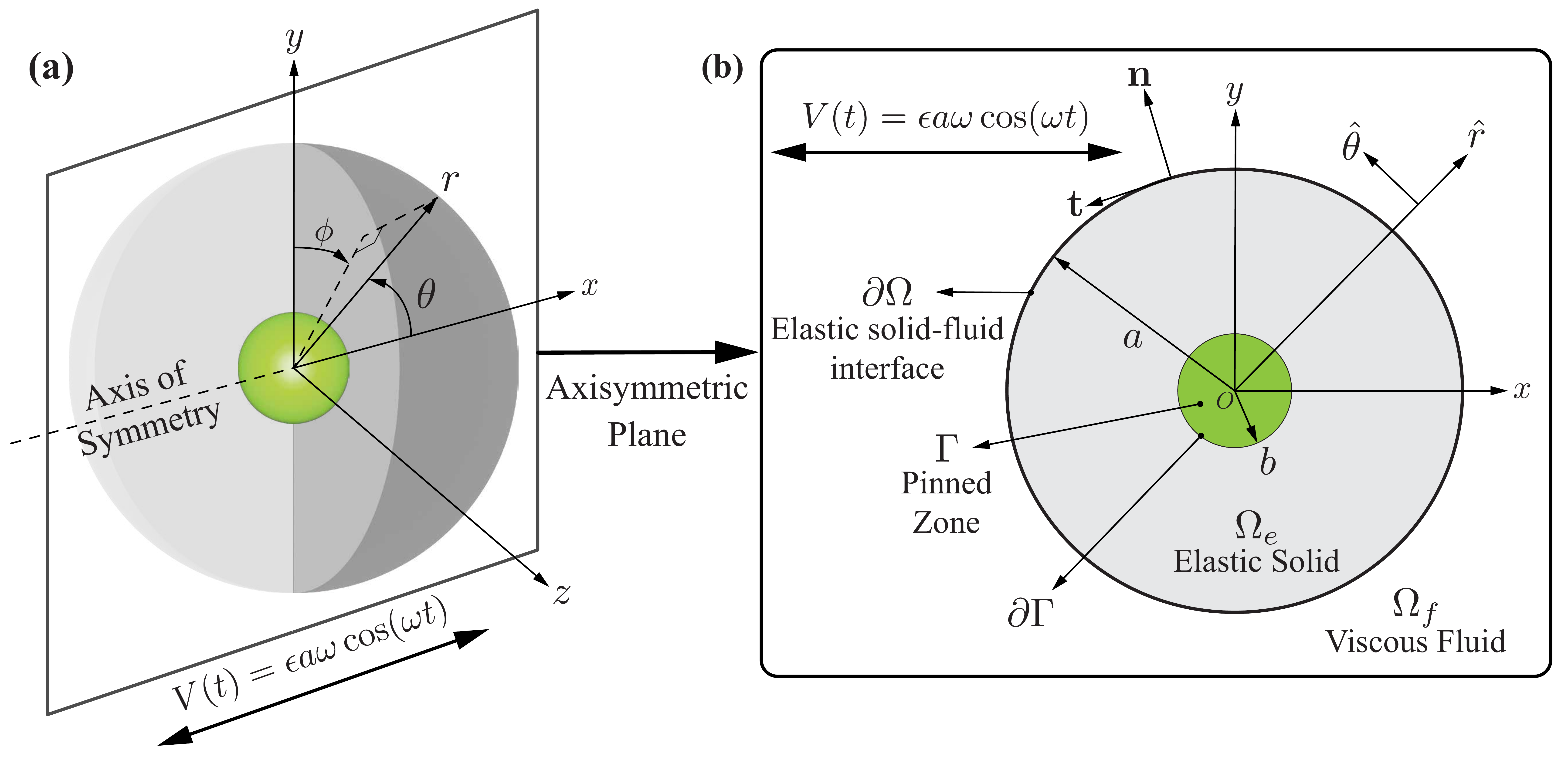}
\caption{\label{fig:setup} Problem setup. (a) 3D Elastic solid sphere $\Omega_e$ of radius \(a\) with a rigid inclusion (pinned zone $\Gamma$ of radius $b$), immersed in viscous fluid $\Omega_f$. 
In this study, we deploy a spherical coordinate system where $(r, \theta, \phi)$ are the radial, polar, and azimuthal coordinates. 
The sphere is exposed to an oscillatory flow with far-field velocity $V(t) = \epsilon a \omega \cos (\omega t)$ in the $x$ direction, along the axis of symmetry. (b) 2D axisymmetric cross-section of the elastic sphere.}
\end{figure}

\section{Problem setup and governing equations}\label{sec:setup}

We derive the streaming solution by considering the setup shown in \cref{fig:setup}, where a 3D visco-elastic solid sphere $\Omega_e$ of radius \( a \) is immersed in a viscous fluid $\Omega_f$. 
The fluid oscillates with velocity $V(t) = \epsilon a \omega \cos \omega t$, where $\epsilon$, $\omega$ and $t$ represent the non-dimensional 
amplitude, angular frequency, and time. 
Following our previous setup for a soft 2D cylinder \cite{bhosale2022streaming}, we kinematically enforce zero strain and velocity near the sphere's center by `pinning' the sphere with a rigid inclusion $\Gamma$ of radius $b < a$, the boundary of which is denoted by \( \partial\Gamma \). 
We further denote by \(\partial\Omega\) the boundary between the elastic solid and the viscous fluid.

Both fluid and solid are assumed to be isotropic and incompressible, where the fluid is Newtonian with kinematic viscosity $\nu_f$ and density $\rho_f$, and the solid follows the Kelvin-Voigt viscoelastic model. 
Characteristic of soft biological materials~\cite{bower2009applied}, elastic stresses within the solid are modeled via neo-Hookean hyperelasticity with shear modulus $G$, kinematic viscosity $\nu_e$, and density $\rho_e$. 
However, we will later show that the choice of hyperelastic or viscoelastic model does not affect the theory.

The dynamics in the fluid and solid phases are governed by the incompressible Cauchy momentum equations, non-dimensionalized using the characteristic scales of velocity $V = \epsilon a \omega$, length $L = a$, time $T = 1 / \omega$, and hydrostatic pressure $P = \mu_f V / L = \mu_f \epsilon \omega$
\begin{equation}
    \begin{aligned}
        \label{eqn:gov_eqns_nondim}
        \textrm{Incomp.}
        &\begin{cases}
         \gv{\nabla} \cdot \gv{v} = 0, ~\gv{x}\in \Omega_f \cup \Omega_e 
        \end{cases}\\
        \textrm{Fluid}
        &\begin{cases}        
        \frac{\partial {\gv{v}}}{\partial {t}} + \epsilon 
        ({\gv{v}} \cdot {\bv{\nabla}}) {\gv{v}} = \frac{1}{M^2} 
        \left(-{\nabla}{p} + {\nabla^2} {\gv{v}} \right),~{\gv{x}}\in\Omega_f
        \end{cases}\\
        \textrm{Solid}
        &\begin{cases} 
        \alpha \Ca \left( \frac{\partial {\gv{v}}}{\partial {t}} + \epsilon 
        ({\gv{v}} \cdot {\bv{\nabla}}) {\gv{v}} \right) = \frac{\Ca}{M^2}
        \left( -{\nabla} {p} + \beta {\nabla^2} {\gv{v}} \right) +
        {\bv{\nabla}} \cdot ({\bv{F}} {\bv{F}}^T)\dev 
        ,~{\gv{x}}\in\Omega_e,
        \end{cases}\\
    \end{aligned}
\end{equation}
where $\gv{v}$ and $p$ are the velocity and pressure fields, and \(\bv{F}\) is the deformation gradient tensor, defined as $\bv{F} = \bv{I} + \bv{\nabla} \gv{u}$, where \(\bv{I}\) is the identity, $\gv{u} = \gv{x} - \gv{X}$ is the material displacement field, and $\gv{x}$, $\gv{X}$ are the position of a material point after deformation and at rest, respectively. 
The prime symbol \(\dev\) on a tensor denotes its deviatoric.  
The key non-dimensional parameters within this system are the scaled oscillation amplitude \(\epsilon \), Womersley number $M = a \sqrt{\rho_f \omega / \mu_f}$, Cauchy number $\Ca = \epsilon \rho_f a^2 \omega^2 / G$, density ratio \(\alpha = \rho_e / \rho_f\), and viscosity ratio \( \beta = \mu_e / \mu_f\). 
Physically, the Womersley number ($M$) represents the ratio of inertial to viscous forces, and the Cauchy number ($\Ca$) represents the ratio of inertial to elastic forces. 
Therefore, a higher $M$ corresponds to stronger dominance of inertia in the fluid environment, and higher $\Ca$ values correspond to increasingly soft bodies. We then impose a set of boundary conditions upon the governing equations, consistent with \citet{lane1955acoustical}
\begin{align*}
    \textrm{Pinned zone}
        &\begin{cases}
            \tag{2.2}
            \label{eqn:bcs_pinned}
            \gv{u} = 0, \gv{v} = 0,~\gv{x}\in \Gamma
        \end{cases}\\
    \tag{2.3}
    \label{eqn:bcs_velocity}
    \textrm{Interface velocity}
        &\begin{cases}
            \gv{v}_e = \gv{v}_f,~\gv{x}\in\partial\Omega        
        \end{cases}\\
    \tag{2.4}
    \label{eqn:bcs_stresses}
    \textrm{Interface stresses}
        &\begin{cases}
            \bv{\sigma}_{f} = -p\bv{I} +  (\bv{\nabla} \gv{v} + \bv{\nabla} \gv{v}^T), ~\gv{x}\in\Omega_f \\
            \bv{\sigma}_{e} = -p\bv{I} + \beta (\bv{\nabla} \gv{v} + \bv{\nabla} \gv{v}^T)
                + \frac{M^2}{\Ca}(\bv{F} \bv{F}^T)\dev,
            ~\gv{x}\in\Omega_e, \\
            \gv{n} \cdot \bv{\sigma}_e \cdot \gv{n} = 
            \gv{n} \cdot \bv{\sigma}_f \cdot \gv{n},~\gv{x}\in\partial\Omega\\
            \gv{n} \cdot \bv{\sigma}_e \cdot \gv{t} = 
            \gv{n} \cdot \bv{\sigma}_f \cdot \gv{t},~\gv{x}\in\partial\Omega\\
        \end{cases}\\
    \tag{2.5}
    \label{eqn:bcs_farfield}
    \textrm{Far-field}
        &\begin{cases}
            \gv{v}(|\gv{x}| \to \infty) =  \cos t ~\hat{i},~\gv{x}\in \Omega_{f},
        \end{cases}
\end{align*}
where \cref{eqn:bcs_pinned} is the pinned-zone rigidity constraint,~\cref{eqn:bcs_velocity} is the no-slip boundary condition between and solid and fluid phases,~\cref{eqn:bcs_stresses} dictates stress continuity, and~\cref{eqn:bcs_farfield} is the far-field flow velocity. 
We use subscripts $e$ and $f$ to denote elastic and fluid phases, respectively, wherever ambiguity may arise. Next, we identify ranges of relevant parameters and solve~\cref{eqn:gov_eqns_nondim} via perturbation theory.

\section{Perturbation series solution}\label{sec:derivation}
In viscous streaming applications, typically we have small non-dimensional oscillation amplitudes $\epsilon \ll {1}$~\cite{wang1965flow,bertelsen1973nonlinear,lutz2005microscopic}, density ratio $\alpha $ and viscosity ratio $\beta$ of $\order{1}$, and Womersley number $M \geq \order{1}$ \cite{marmottant2004bubble,lutz2006hydrodynamic}. 
For the Cauchy number $\Ca$, we apply the same treatment as \citet{bhosale2022streaming}, where we use $\Ca = 0$ for a rigid body and $\Ca = \kappa \epsilon$ with $\kappa = \order{1}$ for elastic bodies. 
The latter assumption implies that $\Ca \ll 1$, which physically means that the body is weakly elastic. 
We make this assumption for two reasons. 
First, we choose $\Ca$ to be small to simplify the treatment of hyperelastic materials, whose non-linearities become mathematically challenging for $\Ca \geq \order{1}$. 
Second, matching $\Ca$ with $\epsilon$ simplifies the asymptotic expansion, while preserving the practical generality of the results (for details, see supplementary material \S 2 of \citet{bhosale2022streaming}).

We then seek a perturbation series solution of Eqs. (\ref{eqn:gov_eqns_nondim}) by asymptotically expanding all relevant fields in powers of $\epsilon$. 
Our derivation closely follows the approach taken by \cite{bhosale2022streaming} for 2D elastic cylinders, while augmenting it to encompass 3D settings.
Then, the zeroth order solution reduces to a rigid sphere in a purely oscillatory flow governed by the unsteady Stokes equation~\cite{lane1955acoustical}. 
The first order solution is subsequently derived in two stages. 
First, we obtain the deformation of the elastic body resulting from the flow field at zeroth order. 
Second, we incorporate the elastic feedback into the streaming solution by using the obtained body deformations as boundary conditions for the flow at $\order{\epsilon}$. The steps are outlined below, with details listed in the supplementary material.

We start by perturbing to \( \order{\epsilon}\) all physical quantities \( q\), which include \(\gv{v}\), \(\gv{u}\), \(p\),  \( \Omega \), \(\gv{n}\),  \(\gv{t}\), as
\begin{equation}
    \label{eqn:perturb}
    q \sim q_0 + \epsilon q_1 + \order{\epsilon^2}
\end{equation}
and substitute them into the governing equations Eqs.~(\ref{eqn:gov_eqns_nondim}) and boundary conditions Eqs.~(\ref{eqn:bcs_pinned})--(\ref{eqn:bcs_farfield}). 
Steps are explicitly reported in the supplementary material (Eq.1.17--1.26), where subscripts (0, 1, ...) refer to the solution order. 
Next, we adopt the geometrically convenient spherical coordinate system $(r, \theta, \phi)$, with radial coordinate $r$, polar angle $\theta$, azimuthal angle $\phi$, and origin at the center of the sphere. 
The horizontal axis direction $\gv{i}$ corresponds to $\theta = 0$. 

\subsection{Zeroth order $\order{1}$ solution}

The governing equations and boundary conditions in the solid phase at zeroth order $\order{1}$ simplify to
\begin{equation}
\label{eqn:gov_eqns_nondim_solid_0}
    \bv{\nabla} \cdot ((\bv{I} + \bv{\nabla} \bv{u}_0) (\bv{I} + \bv{\nabla} \bv{u}_0)^T)\dev = 0,
    ~~ r \leq 1;~~~
    \gv{u}_0|_{r = \zeta} = 0
\end{equation}
where $\zeta = b / a$ is the non-dimensional radius of the pinned zone.
Since at this order $\Ca = \kappa \epsilon = 0$, the solution of \cref{eqn:gov_eqns_nondim_solid_0} physically corresponds to a fixed, rigid sphere with
\begin{equation}
\label{eqn:solid_0}
    \partial\Omega_0 = r = 1;~~
    \gv{u}_0 = 0,~~ 
    \gv{v}_{0} = \frac{\partial \gv{u}_0}{dt} = 0 ,~~ r \leq 1.
\end{equation}
Thus the fluid phase governing equations and boundary conditions reduce to
\begin{equation}
\label{eqn:gov_eqns_nondim_fluid_0}
    \begin{aligned}
        M^2 \frac{\partial \bv{\nabla}^2 \gv{\varphi}_0}{\partial t} &= \bv{\nabla}^4 \gv{\varphi}_{0}~~~~ r \geq 1 \\
        v_{0,r}|_{r = 1} = \frac{1}{r \sin\theta} \frac{\partial (\varphi_0 \sin\theta)}{\partial \theta}\biggr|_{r = 1} &= 0;~
        v_{0, \theta}|_{r = 1} = -\frac{1}{r} \frac{\partial (r \varphi_0)}{\partial r}\biggr|_{r = 1} &= 0 \\
        v_{0, r}|_{r \to \infty} = \cos \theta ~ \cos t &;~
        v_{0, \theta}|_{r \to \infty} = -\sin \theta ~ \cos t,
    \end{aligned}
\end{equation}
where $\gv{\varphi} = \varphi \gv{\hat{\phi}}$ is the vector potential defined as $\gv{v} = \bv{\nabla} \times \gv{\varphi}$, with $\gv{\hat{\phi}}$ being the unit vector in the azimuthal direction. 
We note that the bold font $\bv{\nabla}^2$ refers to the vector Laplacian operator, which is distinct from the scalar Laplacian operator in spherical coordinates. 
This system (Eqs.~(\ref{eqn:solid_0}), (\ref{eqn:gov_eqns_nondim_fluid_0})) defines a rigid sphere immersed in an oscillating unsteady Stokes flow, which has an exact analytical solution \cite{lane1955acoustical}
\begin{equation}
    \label{eqn:soln_fluid_0}
    \varphi_0 = -\frac{\sin\theta}{4} \left( 3\frac{h_1(mr)}{mh_0(m)} - r - \frac{h_2(m)}{r^2 h_0(m)} \right) e^{-it} + c.c. ,~r \geq 1
\end{equation}
where $i = \sqrt{-1}$, and $m = \sqrt{i} M$. 
Here, $h_n$ and $c.c.$ refer to the $n^{\textrm{th}}$ order spherical Hankel function of the first kind and complex conjugate, respectively. 
As observed in \citet{lane1955acoustical}, the zeroth order vector potential field $\gv{\varphi}_0$ in the fluid phase is purely oscillatory in time, and thus no steady streaming appears at this order. 
Moreover, the flow at $\order{1}$ is unaffected by elasticity.

\subsection{First order $\order{\epsilon}$ solution}

We then proceed to the next order approximation $\order{\epsilon}$, where we expect time-independent steady streaming to emerge \cite{lane1955acoustical}. 
At $\order{\epsilon}$, the solid governing equations reduce (supplementary material, Eqs.~1.38-1.42, 1.49) to the homogeneous biharmonic equation
\begin{equation}
    \label{eqn:gov_eqns_nondim_solid_1}
    \bv{\nabla}^4 \gv{\varphi}_{e,1} = 0,~~~~\gv{x}\in\Omega_{e}
\end{equation}
where we have defined the strain function $\gv{\varphi}_{e} = \varphi_e \hat{\gv{\phi}}$ similar to the fluid phase, so that the displacement field is $\gv{u} = \gv{\nabla} \times \gv{\varphi}_{e}$. 
\Cref{eqn:gov_eqns_nondim_solid_1} demonstrates how the specific choice of solid elasticity model used at \(\order{\epsilon}\) becomes irrelevant, since all nonlinear stress-strain responses drop out as a result of linearization (supplementary material, Eqs. ~1.38--1.42). \Cref{eqn:gov_eqns_nondim_solid_1} is complemented by the Dirichlet boundary conditions at the pinned zone interface
\begin{equation}
    \label{eqn:pinned_solid_bc_1}
    u_{1, r} = \left. \frac{1}{r \sin\theta} \frac{\partial (\varphi_{e,1} \sin\theta)}{\partial \theta} \right|_{r = \zeta} = 0;~~~~
    u_{1, \theta} = \left. -\frac{1}{r} \frac{\partial (r \varphi_{e,1})}{\partial r} \right|_{r = \zeta} = 0.
\end{equation}
Additionally, in accordance with~\cref{eqn:bcs_stresses}, the \(\order{1}\) flow solution exerts interfacial stresses on the solid, which at \( \order{\epsilon}\) is no longer rigid but instead deformable. This results in the following boundary conditions for the radial and tangential stresses at the interface $\partial \Omega_0$
\begin{equation}
    \begin{aligned}
        \label{eqn:solid_fluid_deformation_bc_1}
            \frac{M^2}{\kappa} \left.\frac{\partial u_{1,r}}{\partial r}\right|_{r = 1} 
            &= \left.\frac{\partial v_{0,r}}{\partial r}\right|_{r = 1}\\
            \frac{M^2}{\kappa} 
            \left. \left( 
            \frac{1}{r} \frac{\partial u_{1,r}}{\partial \theta} 
            + \frac{\partial u_{1,\theta}}{\partial r} 
            - \frac{u_{1,\theta}}{r}
            \right) \right|_{r = 1} &=
            \left. \left( \frac{1}{r}\frac{\partial v_{0,r}}{\partial \theta} + \frac{\partial
                v_{0,\theta}}{\partial r} - \frac{v_{0,\theta}}{r} \right) \right|_{r = 1},
    \end{aligned}
\end{equation}
where the LHS corresponds to the elastic stresses in the solid phase \big($\frac{M^2}{\Ca}(\bv{F} \bv{F}^T)\dev$, ~\cref{eqn:bcs_stresses}\big) and the RHS to the viscous stresses in the fluid phase ($\bv{\nabla} \gv{v} + \bv{\nabla} \gv{v}^T$, ~\cref{eqn:bcs_stresses}), both evaluated at the zeroth order interface \( r = 1\). 
The pressure term ($-p\bv{I}$, ~\cref{eqn:bcs_stresses}) cancels out due to pressure continuity at the interface, hence its absence in  \cref{eqn:solid_fluid_deformation_bc_1} (supplementary material, Eq.~1.31).
We point out that the use of \( r = 1\) in \cref{eqn:solid_fluid_deformation_bc_1} is consistent despite the fact that the solid interface deforms at this order. Indeed, as demonstrated in \citet{bhosale2022streaming} and supplementary material Eqs.~1.44--1.46, errors associated with the $r=1$ approximation all appear at the higher order $\order{\epsilon^2}$ and thus do not affect our solution.
In addition, we note that the viscous stress term ($\beta (\bv{\nabla} \gv{v} + \bv{\nabla} \gv{v}^T)$, ~\cref{eqn:bcs_stresses}) is also of higher order $\order{\epsilon^2}$ and thus absent in \cref{eqn:solid_fluid_deformation_bc_1}, implying that the specific choice of viscosity model is irrelevant at $\order{\epsilon}$.
Next, we use the $\order{1}$ flow velocity at the interface, \cref{eqn:gov_eqns_nondim_fluid_0,eqn:soln_fluid_0}, to directly evaluate the RHS of \cref{eqn:solid_fluid_deformation_bc_1}
\begin{equation}
    \begin{aligned}
        \left.\frac{\partial v_{0,r}}{\partial r}\right|_{r = 1} &= 0 \\
            \left.\left( \frac{1}{r}\frac{\partial v_{0,r}}{\partial \theta} + \frac{\partial
                v_{0,\theta}}{\partial r} - \frac{v_{0,\theta}}{r} \right)\right|_{r = 1} &= \sin{\theta} ~ F(m) ~ e^{-it} + c.c.    
    \end{aligned}
    \label{eqn:zero_stress}
\end{equation}
with
\begin{equation} 
    \label{eqn:fm}
    F(m) = -\frac{3m h_{1}(m)}{4h_0(m)}.
\end{equation}
With the boundary conditions Eqs.~(\ref{eqn:pinned_solid_bc_1}) -- (\ref{eqn:fm}) resolved, the homogeneous biharmonic~\cref{eqn:gov_eqns_nondim_solid_1} can be solved exactly to obtain the \( \order{\epsilon}\) solid strain function
\begin{equation}
    \label{eqn:strain_func}
    \varphi_{e,1} = \frac{\kappa}{M^2} \sin{\theta} \left( c_1 r + \frac{c_2}{r^2} +  c_3 r^3 + c_4 \right) F(m) ~ e^{-it} + c.c., 
\end{equation}
where the exact expressions for \(c_1, c_2, c_3, c_4 \) (functions of $\zeta$) are reported in the supplementary material (Eq.~1.56).
The $\order{\epsilon}$ solid displacement field $\gv{u}_1$, both in the bulk $\Omega_e$ and at the boundary $\partial \Omega$, can then be directly obtained from \cref{eqn:strain_func}. This, in turn, kinematically affects the flow at $\order{\epsilon}$ via the interfacial boundary conditions, as we will see. 

The flow governing equation at $\order{\epsilon}$, in vector potential form, reads
\begin{equation}
\label{eqn:gov_eqns_nondim_fluid_1}
    M^2 \frac{\partial \bv{\nabla}^2 \gv{\varphi}_1}{\partial t} + 
    M^2 \left( \left( \gv{v}_0 \cdot \bv{\nabla} \right) \bv{\nabla}^2 \gv{\varphi}_0 \right) -
    M^2 \left( ( \bv{\nabla}^2 \gv{\varphi}_0 \cdot \bv{\nabla} ) \gv{v}_0 \right)
    = \bv{\nabla}^4 \gv{\varphi}_{1},~~~~ r \geq 1.
\end{equation}
We note that the term $M^2 (\bv{\nabla}^2\gv{\varphi}_0 \cdot \bv{\nabla})\gv{v}_0$ in \cref{eqn:gov_eqns_nondim_fluid_1}, which corresponds to vortex stretching, is absent in the rigid sphere streaming derivation of \citet{lane1955acoustical}. By considering this unaccounted term, our work improves upon the existing theory, as demonstrated in Section \ref{sec:num_comp}. 

Next, since we are interested in steady streaming, we consider the time-averaged form of \cref{eqn:gov_eqns_nondim_fluid_1}
\begin{equation}
\begin{aligned}
    \label{eqn:gov_eqns_nondim_fluid_1_steady}
    \bv{\nabla}^4 \langle \gv{\varphi}_{1} \rangle &= 
    M^2
    \underbrace{\langle \left( \gv{v}_0 \cdot \bv{\nabla} \right) \bv{\nabla}^2 \gv{\varphi}_0 - ( \bv{\nabla}^2 \gv{\varphi}_0 \cdot \bv{\nabla} ) \gv{v}_0 \rangle}_{\textrm{RHS}}
    ,~~~~ r \geq 1, \\
\end{aligned}
\end{equation}
where we substitute \cref{eqn:soln_fluid_0} into the RHS to yield
\begin{equation}
\begin{aligned}
    \bv{\nabla}^4 \langle \gv{\varphi}_{1} \rangle &= 
    \sin 2 \theta ~ \rho(r) \gv{\hat{{\phi}}}
    ,~~~~ r \geq 1, \\
    \rho(r) &= \frac{1}{16r^4} ~
    \left( r^3J^{(3)} + r^2 J^{(2)}-6r J^{(1)} + 6J \right)J^* + c.c. \\
    J(r) &= 3\frac{h_1(mr)}{mh_0(m)} - r - \frac{h_2(m)}{r^2 h_0(m)}~.
\end{aligned}
\label{eqn:rey_stress}
\end{equation}
Here, $J$ is the radially dependent term of \cref{eqn:soln_fluid_0}, with $J^{(n)}$ and $J^*$ being its $n^{\text{th}}$ derivative and complex conjugate, respectively. Solving this inhomogeneous biharmonic equation requires four independent boundary conditions. The first two are the radial and tangential, time-averaged, far-field velocity
\begin{equation}
    \begin{aligned}
        \label{eqn:fluid_farfield_bc_1}
        \left.\frac{1}{r \sin\theta} \frac{\partial (\langle \varphi_1 \rangle \sin\theta)}{\partial \theta}\right|_{r \to \infty} =
        \left.\frac{1}{r} \frac{\partial (r \langle \varphi_1 \rangle)}{\partial r}\right|_{r \to \infty} = 0.
    \end{aligned}
\end{equation}
Next, we recall the no-slip boundary condition of \cref{eqn:bcs_velocity} that needs to be enforced at the $\order{\epsilon}$ accurate solid--fluid interface
\begin{equation}
    \label{eqn:solid_fluid_bc_1_pre}
    \left. \gv{v}_{e} \right|_{\partial \Omega} = 
    \left. \gv{v}_{e} \right|_{r = 1 + \epsilon u_{1, r}} + \order{\epsilon^2} = 
    \left. \gv{v}_{f} \right|_{\partial \Omega}
    =\left. \gv{v}_{f} \right|_{r = 1 + \epsilon u_{1, r}} + \order{\epsilon^2}.
\end{equation}
We note that the sphere interface at $\order{\epsilon}$ deforms as $r' = 1 + \epsilon u_{1, r}$, where $u_{1, r}$ is the radial component of the $\order{\epsilon}$ accurate displacement field obtained by taking the curl of the strain function $\gv{u}_1 = \bv{\nabla} \times \gv{\varphi}_{e,1}$. 
Similar to \citet{bhosale2022streaming}, we enforce the no-slip boundary condition in \cref{eqn:solid_fluid_bc_1_pre} by deploying the technique presented in \citet{longuet1998viscous}, where $\left. \gv{v}_{f} \right|_{r = r'}$ is Taylor expanded about $r = 1$ (supplementary material, Eqs.~1.64--1.66)
\begin{equation}
    \begin{aligned}
    \label{eqn:solid_fluid_bc_1_expand}
    \left. \gv{v}_{f} \right|_{r = 1 + \epsilon u_{1,r}} &= \left. \left( \gv{v}_{f,1} + \epsilon \frac{\partial \gv{v}_{f,0}}{\partial r} u_{1,r} \right) \right|_{r = 1} + \order{\epsilon^2}.
    \end{aligned}
\end{equation}
The boundary solid velocity $\left. \gv{v}_{e} \right|_{r = 1 + \epsilon u_{1, r}}$ (LHS of~\cref{eqn:solid_fluid_bc_1_pre}) can be instead computed to $\order{\epsilon}$ accuracy as $\left. \partial u_{1,r} / \partial t \right|_{r = 1}$ (supplementary material, Eq. 1.63). 
We note that both $u_{1,r}$ and $\partial \gv{v}_{f,0} / \partial r$ are known from \cref{eqn:strain_func,eqn:soln_fluid_0}. Thus, the $\order{\epsilon}$ flow velocity $\gv{v}_{f,1}$ at $r = 1$, denoted by $\gv{v}_{1}$ henceforth, can be obtained by substituting \cref{eqn:solid_fluid_bc_1_expand} into \cref{eqn:solid_fluid_bc_1_pre} (supplementary material, Eqs.~1.62--1.67). Time averaging then yields the remaining two boundary conditions for \cref{eqn:rey_stress}
\begin{equation}
    \begin{aligned}
        \label{eqn:solid_fluid_bc_1}
        \left. \langle v_{1, r} \rangle \right|_{r = 1} &= 
        \left. \frac{1}{r \sin\theta} \frac{\partial (\langle \varphi_1 \rangle \sin\theta)}{\partial \theta}\right|_{r = 1} = 0~\\
        \left. \langle v_{1, \theta} \rangle \right|_{r = 1} &= -\left.\frac{1}{r}\frac{\partial (r\langle \varphi_1 \rangle)}{\partial r}\right|_{r = 1} = -\frac{\kappa}{M^2} \sin 2 \theta ~ G_1(\zeta) F(m) F^*(m)
    \end{aligned}
\end{equation}
with
\begin{equation} 
    \label{eqn:g1}
    G_1(\zeta) = \frac{(\zeta - 1)^2(4\zeta^2 + 7\zeta + 4)}{3\zeta(2\zeta^3 + 4\zeta^2 + 6\zeta + 3)}.
\end{equation}
\Cref{eqn:solid_fluid_bc_1} physically implies a rectified tangential slip velocity ($\langle v_{1, \theta} \rangle|_{r = 1} \neq 0$) in the fluid phase at the zeroth-order fixed interface $r = 1$. 
This slip velocity captures the effect of body elastic deformation ($\langle v_{1, \theta} \rangle|_{r = 1} = 0$ for rigid bodies) by equivalently modifying the fluid Reynolds stresses ($\sin 2\theta \rho(r)$ in~\cref{eqn:rey_stress}), thus impacting the resulting streaming flow.
We further remark that, in contrast to rigid body streaming, such modification is accessible even in the Stokes limit as it is independent of the Navier-Stokes nonlinear inertial advection, a conclusion similarly drawn in our previous work on 2D soft cylinder streaming~\cite{bhosale2022streaming}. 
This phenomenon shares characteristics with the artificial mixed-mode streaming of pulsating bubbles \cite{longuet1998viscous,spelman2017arbitrary}, whereas the streaming process derived here arises spontaneously from the coupling between viscous fluid and elastic solid. 

\begin{figure}
\centering
\includegraphics[width=\linewidth]{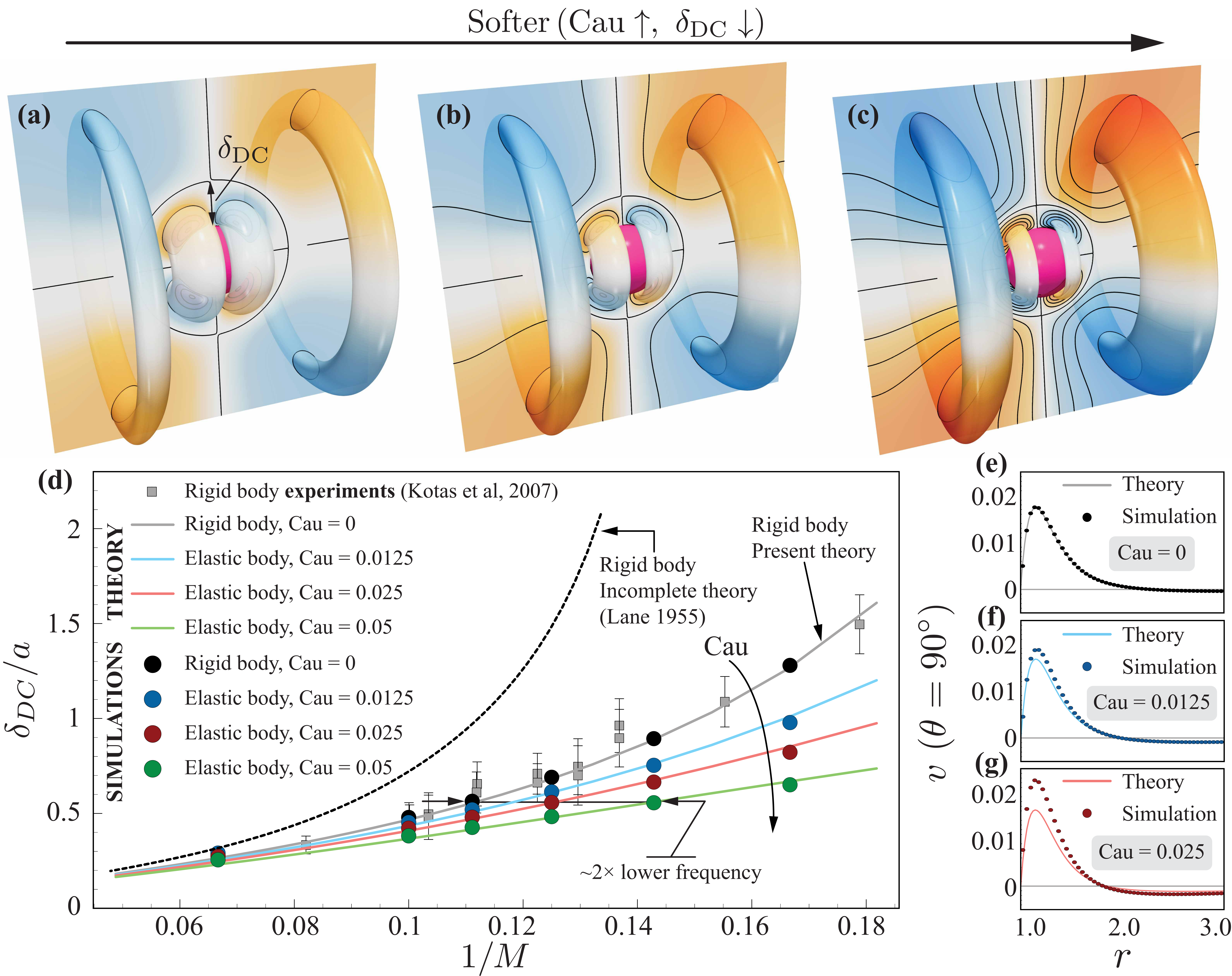}
\caption{\label{fig:DClayer} \textbf{Elastic sphere and streaming flow response.}
\textbf{(a-c)} 3D time-averaged Lagrangian (i.e. Stokes-drift corrected, supplementary material \S{3}) Stokes streamfunction depicting the streaming response at $M = 6$ with increasing softness $\Ca$. (a) Rigid limit $\Ca = 0$, (b) $\Ca = 0.025$, and (c) $\Ca = 0.05$. Note that blue/orange represent clockwise/counterclockwise rotating regions. 
The non-dimensional radius of the pinned zone is set at $\zeta = 0.4$ throughout the study, to maintain the tangential slip velocity magnitude (Eq. \ref{eqn:solid_fluid_bc_1}) at $\order{1}$, consistent with the asymptotic analysis. 
Lowering the pinned zone radius rapidly increases the prefactor $G_1(\zeta)$ (Eq.~\ref{eqn:g1}), which results in a slip velocity much greater than $\order{1}$, thereby weakening the asymptotic assumption. The opposite holds true for large $\zeta$, rendering $\zeta = 0.4$ a robust compromise.
The effect of pinned zone radius on streaming flow is detailed in Section \S{4} of the supplementary material.
\textbf{(d)} Normalized DC layer thickness ($\delta_{DC} / a$) vs. inverse Womersley number ($1 / M$) from theory and simulations, for varying body elasticity $\Ca$. 
Incomplete theory (dashed line) \cite{lane1955acoustical} and experimental results (grey squares) \cite{Kotas:2007} in the rigidity limit are plotted for reference. 
We characterize streaming via the thickness of the DC layer, which refers to the innermost recirculation zone, for its utility in trapping, filtration and chemical mixing, and because of its robust nature. 
\textbf{(e-g)} Radial decay of velocity magnitude along $\theta = 90^{\circ}$ from theory and simulations at $M = 6$, with increasing softness $\Ca$. (e) Rigid limit $\Ca = 0$, (f) $\Ca = 0.0125$, and (g) $\Ca = 0.025$. Simulations are performed using a vortex-method based formulation \cite{gazzola2011simulations,bhosale2021remeshed,pyaxisymflow2023} that solves the vorticity form of the Navier--Stokes equations in an axisymmetric cylindrical coordinate system. 
Within this framework, fluid and solid are modeled to be density matched ($\rho_f = \rho_e = 1$). The rigid sphere and pinned zone are modeled as Brinkman solids, and the elastic phase as a viscoelastic Kelvin-Voigt solid with shear modulus $G$. Soft body deformations are tracked using an inverse-map technique \cite{bhosale2021remeshed,kamrin2009eulerian,kamrin2012reference}. The far-field velocity is $V(t) = V_0 \cos \omega t$ with characteristic velocity $V_0 = \epsilon a \omega$, where $\epsilon = 0.1$, $a = 0.1$, and $\omega = 32 \pi$. The fluid dynamic viscosity $\mu_f$ and elastic body's shear modulus $G$ are determined based on the Womersley number $M = a \sqrt{\rho_f \omega / \mu_f}$ and Cauchy number $\Ca = \epsilon \rho_f a^2 \omega^2 / G$. Additional simulation parameters include: domain $[0, 1]\times[0, 0.5]$, uniform grid spacing $h = 1 / 1024$, Brinkman penalization factor $\lambda = 1e6$, mollification length $\epsilon_{moll} = 2h$, Courant–Friedrichs–Lewy number CFL = $0.1$. For details on these parameters, refer to \cite{gazzola2011simulations,bhosale2021remeshed,parthasarathy2022elastic,pyaxisymflow2023}.} 
\end{figure}

Given the steady flow of Eq.~(\ref{eqn:rey_stress})
and boundary conditions of \cref{eqn:fluid_farfield_bc_1,eqn:solid_fluid_bc_1}, the streaming solution can finally be written as
\begin{equation}
    \label{eqn:soln_fluid_1_steady}
    \langle \varphi_1 \rangle = \sin 2 \theta ~ \left[ \Theta(r) + \Lambda(r) \right],
\end{equation}
where $\Theta(r)$ is the rectified rigid-body solution
\begin{equation}
    \begin{aligned}
    \label{eqn:soln_fluid_1_rigid}
        \Theta(r) &= -\frac{r^{4}}{70} \int_{r}^{\infty} \frac{\rho(\tau)}{\tau} ~ \mathrm{d} \tau + \frac{r^{2}}{30} \int_{r}^{\infty} \tau \rho(\tau) ~ \mathrm{d} \tau \\
        &+ \frac{1}{r}\left( \frac{1}{30} \int_{1}^{r} \tau^{4} \rho(\tau) ~ \mathrm{d} \tau + \frac{1}{20} \int_{1}^{\infty} \frac{\rho(\tau)}{\tau} ~ \mathrm{d} \tau - \frac{1}{12} \int_{1}^{\infty} \tau \rho(\tau) ~ \mathrm{d} \tau\right) \\
        &+ \frac{1}{r^{3}}\left(-\frac{1}{70} \int_{1}^{r} \tau^{6} \rho(\tau) ~ \mathrm{d} \tau - \frac{1}{28} \int_{1}^{\infty} \frac{\rho(\tau)}{\tau} ~ \mathrm{d} \tau \right. \left. + \frac{1}{20} \int_{1}^{\infty} \tau \rho(\tau) ~ \mathrm{d} \tau\right)
    \end{aligned}
\end{equation}
and $\Lambda(r)$ is the new elastic modification
\begin{equation}
    \label{eqn:soln_fluid_1_elastic}
    \Lambda(r) = 0.5 \frac{\kappa}{M^2} ~ G_1(\zeta) F(m) F^*(m) \left( \frac{1}{r} - \frac{1}{r^3} \right)
\end{equation}
with $G_1(\zeta)$ and $F(m)$ given in~\cref{eqn:g1} and  \cref{eqn:fm}, respectively. We note that while \cref{eqn:soln_fluid_1_rigid} is of the same form as \citet{lane1955acoustical}'s solution, the explicit expression of $\rho(r)$ is different (Eq.~\ref{eqn:rey_stress}) because of the vortex stretching term of \cref{eqn:gov_eqns_nondim_fluid_1}.
This concludes our theoretical analysis.

\section{Numerical validation and extension to complex bodies}\label{sec:num_comp}
Next, we compare our theory against known experimental and analytical results in the rigidity limit \cite{Kotas:2007,lane1955acoustical}, as well as direct numerical simulations performed using an axisymmetric vortex-method based formulation \cite{gazzola2012c,bhosale2021remeshed,pyaxisymflow2023} (see also the caption of Fig. 2). 
The Stokes streamfunction pattern for a rigid sphere (\( \Ca = 0\)) oscillating at $M \approx 6$ is shown in \cref{fig:DClayer}a    . 
We highlight the two-fold symmetry on top of the axisymmetry, and the presence of a well-defined direct circulation (DC) layer of thickness $\delta_{DC}$. 
This characteristic flow configuration, as well as the divergence of the DC layer thickness ($\delta_{DC}\rightarrow \infty$) with increasing $1/M$, is consistent with \citet{lane1955acoustical}.
This qualitative behavior is recovered by our theory at \( \Ca = 0\) (grey line in \cref{fig:DClayer}d), and by simulations (black dots in \cref{fig:DClayer}d). 
However, we note the significant quantitative difference between the results from \citet{lane1955acoustical} (black dashed line in \cref{fig:DClayer}d) and our simulations/theory, which are instead found to be in close agreement with experimental results (grey dots) by \citet{Kotas:2007} (grey squares). 
The additional accuracy of our theory directly stems from including the vortex stretching term of \cref{eqn:gov_eqns_nondim_fluid_1_steady}, as previously discussed.
Next, as we enable solid compliance (\( \Ca > 0\)), we observe that the two-fold symmetry is preserved ($\sin 2 \theta$ in \cref{eqn:soln_fluid_1_steady}) while $\delta_{DC}$ contracts due to the elastic modification term $\Lambda \neq 0$, in agreement with numerical simulations across a range of $\Ca$ (\cref{fig:DClayer}(b-d)). 
These observations are consistent with our previous work on streaming for a 2D soft cylinder \cite{bhosale2022streaming}, and thus a similar, intuitive explanation exists. 
The flow receives feedback deformation velocities on account of the deformable sphere surface, which acts as an additional source of inertia.
Since the Womersley number (M) is the ratio between inertial and viscous forces, this is equivalent to rigid body streaming with a larger $M$, hence the decrease of DC layer thickness with increasing elasticity. 
This implies that an elastic body can access the streaming flow configurations of its rigid counterpart with significantly lower oscillation frequencies. 
Such frequency reduction is shown in \cref{fig:DClayer}d where, for example at $\Ca = 0.05$, the same DC layer thickness is achieved at $\sim 2 \times$ lower frequency.  
Similar to rigid objects, the $\delta_{DC}$ of a soft sphere still diverges with decreasing $M$, albeit at lower values, since the elastic modification $\Lambda(r)$ does not alter the asymptotic behavior of the rigid contribution $\Theta(r)$ (see supplementary material, Section 6 for details). 
We further note that as $\Ca$ increases and $M$ decreases, the deviation between simulations and theory grows. This follows from the fact that as $\kappa / M^2$ in \cref{eqn:solid_fluid_bc_1} increases, the tangential slip velocity assumed to be of $\order{\epsilon}$ can exceed $\order{1}$, eventually leading to the breakdown of the asymptotic analysis.
We conclude our validation by showing close agreement between theoretical and simulated radially-varying, time-averaged velocities \( \langle v \rangle \) at \(\theta = 90^{\circ} \) (\cref{fig:DClayer}(e--g)). 
For a detailed analysis concerning the effect of inertia ($M$) and elasticity (Cau) on velocity magnitudes (flow strength), refer to Section 4 of the supplementary material.

\begin{figure}
\centering
\includegraphics[width=\linewidth]{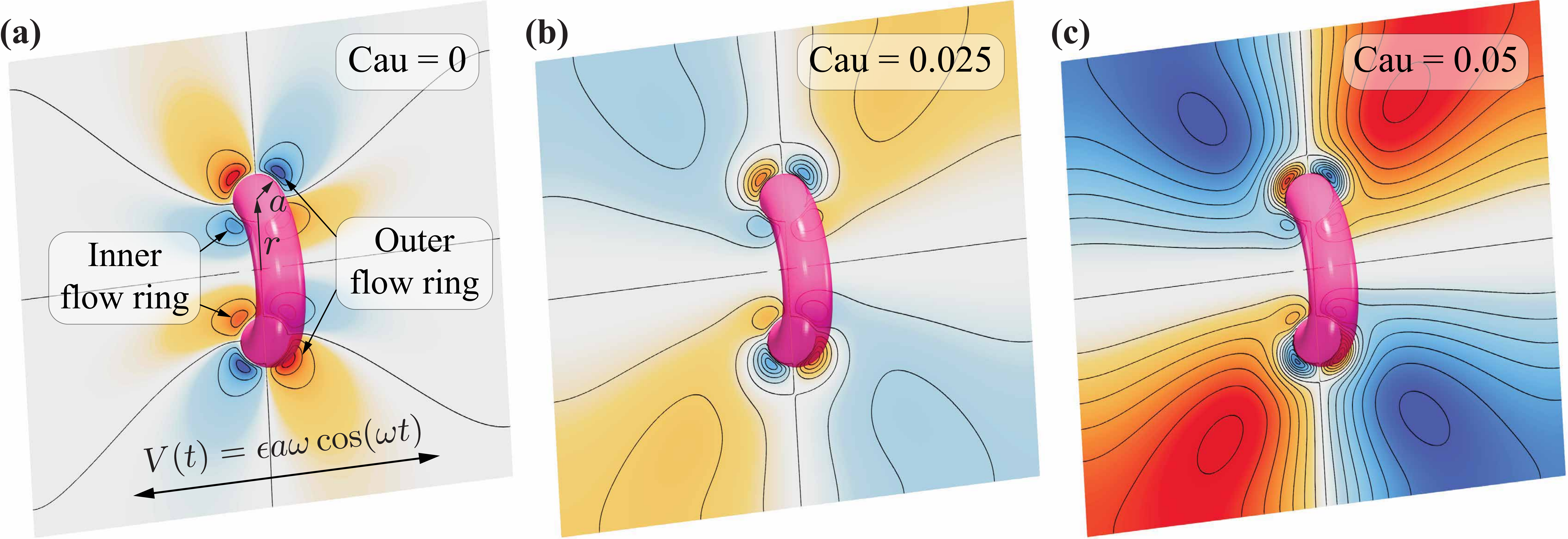}
\caption{\label{fig:torus} \textbf{Extension to complex bodies}. 
Here we consider compliance-induced streaming in a soft torus, a complex shape entailing multiple curvatures and distinct topology relative to the sphere.
Numerically simulated time-averaged Eulerian flow topologies for a torus of core radius $r$ and cross-sectional radius $a = r/3$, at $M \approx 4$, with varying body elasticity $\Ca$. \textbf{(a)} Rigid limit $\Ca = 0$, \textbf{(b)} $\Ca = 0.025$, and \textbf{(c)} $\Ca = 0.05$. The viscous fluid oscillates with velocity $V(t) = \epsilon a \omega \cos \omega t$. The torus is `pinned' at the center of its circular cross-section by a rigid toroidal inclusion of radius $0.4a$. All other physical and simulation parameters are consistent with sphere streaming (see captions of \cref{fig:DClayer} for details).}
\end{figure}

Finally, we demonstrate how gained theoretical insights translate to 3D geometries characterized by multiple curvatures and distinct topology, illustrated here by means of a torus, a shape of interest due to its microfluidic properties \cite{chan2021three} and recent bioengineered demonstrations \cite{dou2023musclering}. 
\Cref{fig:torus}a presents the streaming flow generated for a rigid torus immersed in an oscillatory flow field at $M \approx 4$. 
As can be seen, the highlighted recirculating flow features (inner/outer flow rings that may be used for particle manipulation) are weak for practical applications in the rigid limit.
This can be remedied by increasing elasticity ($\Ca$), for which we indeed observe enhanced flow strengths (\cref{fig:torus}(b,c)).
Finally, we highlight that to obtain a flow topology similar to \cref{fig:torus}c, but with a rigid torus, oscillation frequency $\sim 4\times$ higher are necessary (supplementary, section \S{7}), in conformity with the intuition gained via the soft sphere streaming analysis.

\section{Conclusion}\label{sec:conclusion}
In summary, this study improves existing three-dimensional rigid sphere streaming theory, expands it to the case of elastic materials, and further corroborates it by means of direct numerical simulations. 
Our work reveals, in keeping with our previous work on 2D soft cylinders, an additional streaming mode accessible through material compliance and available even in Stokes flow. 
It further demonstrates how body elasticity strengthens streaming or enables it at significantly lower frequencies relative to rigid bodies.
Finally, we show how theoretical insights extend to geometries other than the sphere, highlighting the practical generality of our theory. 
Overall, these findings advance our fundamental understanding of streaming flows, with potential implications in both biological and engineering domains.

\noindent \textbf{Funding.}
This work was supported by the NSF CAREER Grant No. CBET-1846752 (MG).

\noindent \textbf{Declaration of Interests.} 
The authors report no conflict of interests.

\bibliographystyle{jfm}
\bibliography{cfs_lit.bib}

\end{document}